\documentclass[debug,overfull]{epl}

\usepackage{amsmath,amssymb}

\title{Normal Transport Behavior in Finite One-Dimensional Chaotic Quantum Systems}
\shorttitle{Normal Transport in Chaotic Quantum Systems}
\author{R. Steinigeweg\inst{1}\thanks{Email: \email{rsteinig@uos.de}} \and J. Gemmer\inst{1} \and M. Michel\inst{2}}
\shortauthor{R. Steinigeweg \etal}
\institute{
  \inst{1} Physics Department, University of Osnabr\"uck - Barbarastrasse 7,  49069 Osnabr\"uck, Germany \\
  \inst{2} Institute of Theoretical Physics I, University of Stuttgart - Pfaffenwaldring 57, 70550 Stuttgart, Germany \\
}

\begin{document}

\maketitle

\begin{abstract}
We investigate the transport of energy, magnetization, etc. in several finite
one-dimensional (1D) quantum systems only by solving the corresponding
time-dependent Schr\"odinger equation. We explicitly renounce on
any other transport-analysis technique. Varying model parameters
we find a sharp transition from non-normal to normal transport and a 
transition from integrability to chaos, i.e., from Poissonian to 
Wigner-like level statistics. These transitions always appear in
conjunction with each other. We investigate some rather 
abstract ``design models'' and a (locally perturbed) Heisenberg spin 
chain.
\end{abstract}

The transport behavior of one-dimensional (1D) systems has
intensively been investigated for several decades, as well in the
context of classical mechanics as in the context of quantum
mechanics \cite{mejiamonasterio2005, saito1996, casati1984,
li2004, michel2003, michel2005, saito2003, garrido2001, lepri2003,
castella1995, zotos1997, narozhny1998, rabson2004, fabricius1998,
zotos2003, benz2006}. Nevertheless, the precise conditions under
which normal transport occurs, i.e., under which there
is neither ballistic transport nor localization but normal spatial diffusion, are still not known
\cite{buchanan2005}. 

In the classical domain it seems to be largely accepted that
normal transport (in any dimension) requires the chaotic dynamics
of a non-integrable system whereas non-normal transport is typical
for the regular dynamics of (completely) integrable systems, see
\cite{casati1984}. However, there have also been successful
attempts to observe normal transport in the absence of exponential
instability, the latter being a basic feature of (deterministic)
chaos \cite{li2004}. In the quantum domain there are only very few
examples which can be reliably shown to exhibit normal, diffusive
transport at all \cite{michel2003, michel2005, saito2003}. But,
also in this field, it has been argued that non-normal transport
is related to the macroscopic number of conserved quantities which
characterize integrable systems \cite{castella1995, zotos1997,
narozhny1998, rabson2004}. Moreover, recent numerical computations
for spin chains have led to the assumption that normal transport
might strictly depend on quantum chaos \cite{mejiamonasterio2005,
saito1996}.

Although this assumption is plausible, it has not been proved yet
\cite{zotos2003}. Moreover, almost all computations are either
based on special models of reservoirs which might effect the
transport behavior, or they rely on the Kubo formula. The latter
has originally been derived for field-driven electrical
conductance and its validity for diffusive transport phenomena
such as, e.g., thermal conductance is still under dispute, see
\cite{saito2003, garrido2001, lepri2003, bonetto2000}. We
refer to \cite{gemmer2006-1} where the rather limited validity of 
the Kubo formula for thermal conductance has explicitly been 
demonstrated. Especially considerations which are restricted to the 
analysis of the so-called ``Drude-weight'' are not sufficient to
determine the transport behavior. Thus the main intent of the letter 
at hand is to examine the relation  between transport behavior and 
quantum chaos without modelling, e.g., external heat baths at different
temperatures or using the Kubo formula.

This letter is structured as follows: First of all we briefly
comment on the theory of quantum chaos, mainly on the nearest
neighbor level spacing distribution (NNLSD). Thereafter we
introduce two measures for deviations: i) a measure for the
deviation of a given system from the fully chaotic Wigner NNLSD
($\chi_W$), ii) a measure for the deviation of a given system from
fully diffusive behavior ($D$). Different finite 1D models are
considered and for each model a parameter that drives an
``integrable to chaotic transition'' is varied. The corresponding
Schr\"odinger equations are solved. This allows to plot both
measures ($\chi_W$, $D$) over the respective parameter in order to
reveal correlations. This reveals that, at least in our models,
the ``integrable to chaotic'' induces a ``non-normal to normal
transport'' transition. We investigate some ``design models''
featuring random interactions and a (locally perturbed) Heisenberg
spin chain. We close with a summary and a conclusion.

The theory of quantum chaos is principally concerned with the
level statistics of quantum systems which possess a classical
limit \cite{haake2004, reichl2004, avishai2002, kudo2004,
santos2004, kolovsky2004, finkel2005}. A commonly used statistical
measure is the nearest neighbor level spacing distribution
(NNLSD), $P(s)$, where $P(s) \, ds$ is the probability that the
distance, $s$, between two adjacent eigenvalues lies in the
interval $[s,s + ds]$. Typically, $P(s)$ is well described by a
Wigner distribution $P_W(s)=\pi s/2 \exp(-\pi s^2/4)$, when the
classical limit is chaotic, and by a Poissonian distribution
$P_P(s) = \exp(-s)$, when the classical limit is regular, i.e.,
(completely) integrable \cite{haake2004, reichl2004}.
Interestingly, $P(s)$ can differ from these distributions
\cite{finkel2005}. Two things are crucial for the computation of
the NNLSD: First of all one has to select a subspace consisting of
states from a single symmetry class. Thereafter one has to unfold
the subspace's spectrum, such that the local average of $s$ equals
one ($\bar{s}=1$) everywhere in the spectrum. A detailed
description of the unfolding procedure can be found in
\cite{haake2004, reichl2004}. In order to compare the resulting
NNLSD, given as a normalized histogram with $L$ bins, with the
above distributions we define the measure
\begin{equation}
\chi_W^2 = \sum_{\mu=1}^L \frac{(P_\mu - P_{W,\mu})^2}{P_{W,\mu}} \; , \label{chisquare}
\end{equation}
and $\chi_P^2$, respectively, where $P_\mu$ is the probability
that $s$ lies inside the $\mu$'th bin of the histogram.
$P_{W,\mu}$ and $P_{P,\mu}$ are the probabilities according to
$P_W(s)$ and $P_P(s)$, respectively.

In this letter we consider chain-like quantum systems which may be
described by Hamiltonians of the form
\begin{equation}
\hat{H} =\sum_{\mu=1}^N \hat{h}_\mu + \sum_{\mu=1}^{N-1}
\hat{v}_{\mu,\mu+1} \; , \label{hamiltonian}
\end{equation}
where $\hat{h}_\mu$ denotes the local Hamiltonian of some subunit
$\mu$ and $\hat{v}_{\mu,\mu+1}$ the interaction between
neighboring subunits, $N$ is the total number of subunits. The model
is primarily intended to investigate energy transport in a chain
of coupled quantum systems (like, e.g., molecules) but may also be
viewed as a Hubbard-type model for
particles on a lattice (see below). Furthermore, most
(almost) periodic systems should allow for a description according
to (\ref{hamiltonian}), as will be illustrated below with the
example of a spin chain and is explained in detail in
\cite{gemmer2006-1}. Since we intend to identify the diffusion of,
e.g., energy, we have to introduce a measure for the energy
density or the local energy. We define the local energy operator
at site $\mu$ simply as $\hat{h}_\mu$. Of course, this definition
neglects the energy contained in the interaction and eventually
implies a weak coupling limit.

If the transport behavior of, e.g., energy was perfectly diffusive, then the
local energies $E_\mu(t)=\langle \psi(t)|\hat{h}_\mu
|\psi(t)\rangle$ ($|\psi(t)\rangle$ being the full system's
wavefunction) would obey the following set of equations for, e.g.,
a chain-like system
\begin{eqnarray}
&& \dot{E}_1 \; = \eta \, (E_2 - E_1) \; , \nonumber \\ %
&& \dot{E}_\mu \; = \eta \, \big[(E_{\mu+1}-E_\mu)-(E_\mu-E_{\mu-1})\big]\; , \label{rates} \\ %
&& \dot{E}_N = \eta \, (E_{N-1} - E_N) \; . \nonumber %
\end{eqnarray}
This may be viewed as a discrete form of the diffusion equation
$\dot{\rho}_E=\eta \Delta\rho_E$, $\rho_E$ being the energy
density. Diffusive behavior of any other quantity may be defined
by a respective set of equations.

A suitable measure for the deviation of the (numerically) exact
time evolution of the local energies as obtained from the
Schr\"odinger equation from the fully diffusive dynamics as generated
by (\ref{rates}) is given by
\begin{equation}
D^2 = \frac{1}{N} \frac{1}{5 \tau} \sum_{\mu=1}^N \int_0^{5 \tau}
\big[E_\mu^{\text{normal}}(t) - E_\mu^{\text{exact}}(t)\big]^2 \, dt
\label{dsquare}
\end{equation}
with $\tau = 1/\eta$. In the letter at hand we determine $\eta$
through fitting the dynamics generated by (\ref{rates}) to the
(numerically) exact time evolution of the local energies. For a
possible analytic computation of $\eta$, see, e.g.,
\cite{michel2005}.

\begin{figure}
  \centering
  \includegraphics[width=7cm]{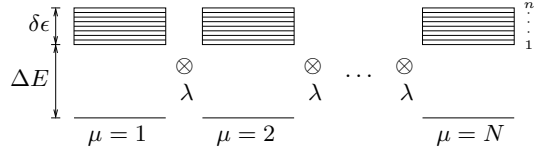}
  \caption{A chain of $N$ identical subunits: Each
    subunit features a nondegenerate ground state, a wide energy gap
    ($\Delta E$) and a comparatively narrow energy band ($\delta
    \epsilon$) which contains $n$, energetic equidistant, states.}
  \label{fig1}
\end{figure}

Let us now investigate two abstract examples (``design models'')
of (\ref{hamiltonian}). The first model consists of $N$ identical
subunits: Each subunit features a nondegenerate ground state, a
wide energy gap ($\Delta E$) and a comparatively narrow energy
band ($\delta \epsilon$) which contains $n$, energetic
equidistant, states, cf.\ Fig.\ \ref{fig1}. The next-neighbor
interaction is defined as
\begin{equation}
\hat{v}_{\mu,\mu+1} = \lambda \sum_{i,j = 1}^n v_{ij} \,
\hat{p}_{\mu,i}^+ \, \hat{p}_{\mu+1,j}^- \, + \, h.c. \; ,
\label{interaction}
\end{equation}
where $h.c.$ is the hermitian conjugate of the previous sum.
$\hat{p}_{\mu,i}^+$ corresponds to an upwards transition of the
$\mu$'th subunit from its ground state to the $i$'th state of its
band and $\hat{p}_{\mu,i}^-$ corresponds to a downwards
transition, respectively. $v_{ij}$ are randomly distributed
complex numbers which are normalized to $\sum_{i,j = 1}^n
|v_{ij}|^2 / n^2 = 1$, such that $\lambda$ sets the total
interaction strength. Since $v_{ij}$ does not change with $\mu$, 
there is no disorder. Obviously, we can restrict our analysis to
the one-excitation subspace (the space where one subsystem is
excited and all others are in their ground state), if the initial
state $\psi(0)$ belongs to this subspace. The subspace's dimension
is $N \cdot n$ and grows linearly rather than exponentially with $N$.

\begin{figure}
  \centering
  \includegraphics[width=14cm]{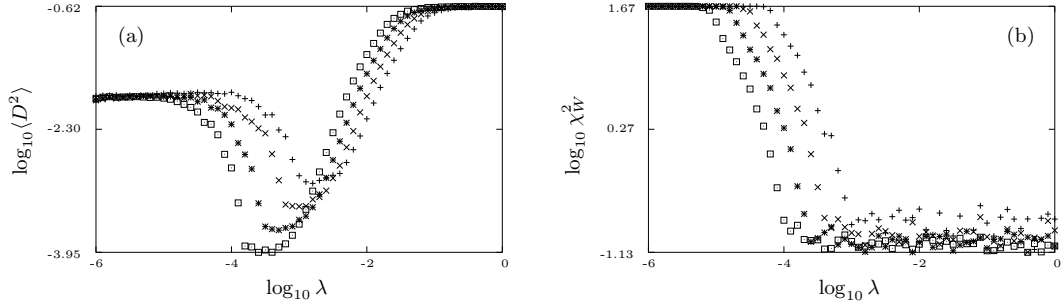}
  \caption{\textbf{(a)} Average deviation of the (numerically) exact
  time evolutions of the local energies from perfectly diffusive
  behavior. \textbf{(b)} Deviation of the level statistics from the
  Wigner distribution for the model pictured in Fig.\ \ref{fig1}.
  The model parameters $N=2$, $\Delta E = 10$ and $\delta \epsilon = 0.5$
  are fixed, but $n$ runs from $125$ ($+$) over $250$ ($\times$) and $500$
  ($*$) till $1000$ ($\boxdot$). Evidently, Wigner-like level statistics
  and normal transport are correlated.}
  \label{fig3}
\end{figure}

The model may be also illustrated as a ``single-particle multi-channel
quantum wire'', as already mentioned above, with random hoppings but 
without disorder. Therefore, unlike the single-particle Hubbard models with 
randomness in, e.g., \cite{ossipov2006}, no localization occurs. Thus
all deviations from normal transport which are discussed below
(see Fig.\ \ref{fig3}a) are deviations towards a ballistic-type 
behavior, not towards localization.

For simplicity, we start with $N = 2$, i.e., only two subunits.
Figure \ref{fig3}a shows $D^2$ versus $\lambda$, averaged for
$200$ adequately restricted random pure initial states $\psi(0)$,
cf.\ (\ref{dsquare}). The model parameters $\Delta E = 10$ and
$\delta \epsilon = 0.5$ are fixed, but $n$ runs from $125$ to
$1000$. Figure \ref{fig3}b shows $\chi_W^2$ versus $\lambda$, that
is, the deviation of the corresponding NNLSD from the Wigner
distribution, cf.\ (\ref{chisquare}). Expectedly, for too small
$\lambda$ ($< 10^{-5}$) the subunits are almost uncoupled, such
that the time evolution of the local energies differs from
(\ref{rates}) and the NNLSD deviates from the Wigner distribution. 
Evidently, (\ref{rates}) does not apply for too large 
$\lambda$ ($> 10^{-2}$), although the NNLSD is clearly Wigner-like. 
Obviously, a Wigner-like NNLSD is not a sufficient condition for normal
transport. Nevertheless, for all $n$ one observation is striking:
The minimum of $D^2$ lies exactly at the position where the
minimum of $\chi_W^2$ is reached. This observation which turns out
to apply to all our models is our main result. Thus, a Wigner-like
NNLSD might be a necessary condition for normal transport.

We have additionally checked chains up to $N = 15$ and $n = 500$.
Furthermore, we have varied details of the model, e.g., the band's
level distribution. We shortly summarize that the results do not
significantly differ from the results of the case $N = 2$. The
interested reader is referred to \cite{michel2005, gemmer2006-1}
where, e.g., the finite size scaling is discussed in detail.

\begin{figure}
  \centering
  \includegraphics[width=7cm]{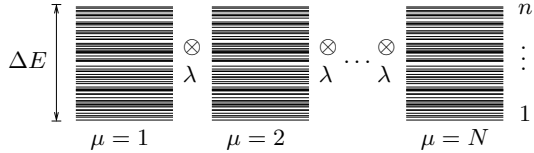}
  \caption{A chain of $N$ identical subunits: Each subunit features
  $n$ states which are randomly, but uniformly distributed within an
  energy interval $\Delta E$.}
  \label{fig4}
\end{figure}

However, since real physical systems do not typically feature the
gapped local spectra of the above model, we consider another
example of (\ref{hamiltonian}). The model is pictured in Fig.\
\ref{fig4} and consists of $N$ identical subunits: Each subunit
features $n$ eigenstates which are randomly, but uniformly
distributed within an energy interval $\Delta E$. As before
$\hat{v}_{\mu,\mu+1}$ is a randomly chosen complex matrix, but now
without restriction to any subspace (no ``particle-number conservation''). Nevertheless,
$\hat{v}_{\mu,\mu+1}$ is supposed to be given by
$\hat{v}_{\mu,\mu+1}$ $= \lambda \, \hat{v}$ with $\text{Tr}
(\hat{v}^2)/n^2 = 1$. Again, for each $\lambda$ we average over an
adequate set of 200 random pure initial states. For the case $N =
2$ the results do not qualitatively differ from the results in
Fig.\ \ref{fig3}a and \ref{fig3}b, e.g., for the model parameters
$n = 60$ and $\Delta E = 10$ the minimum of $D^2$ and $\chi_W^2$
lies at $\lambda = 0.0005$.

What about larger $N$? Since for this system class the dimension of 
the relevant Hilbert space is $n^N$, we are forced to decrease $n$. But for 
smaller $n$ deviations from the fully diffusive behavior typically increase,
cf.\ Fig.\ \ref{fig3}a. Nevertheless, in principle we find the
above result also confirmed for $N = 3$ and $n = 20$. Remarkably,
even for $N = 6$ and $n = 4$ the time evolution of the local
energies is in tolerably good agreement with (\ref{rates}), again
for a parameter regime where the NNLSD is Wigner-like. (We do not
display all those data here, since they essentially look like
Fig.\ \ref{fig3}.)
 
However, for a chain of two level subunits,
that is, for a chain of ``spins'' relaxation and local
fluctuations are indistinguishable on the scale of a single
subunit, i.e., on this scale (\ref{rates}) does not apply. This
changes if the scale is changed, i.e., if various neighboring
spins (including their mutual interactions) are grouped together
to form a subunit as addressed by (\ref{hamiltonian}). However,
since a spin chain with random next neighbor interaction may
always be viewed as a ``mixture'' of different wellknown spin
models, it is more meaningful to consider directly a single spin
model, e.g., the Heisenberg model. Furthermore, other than our
``design models'', those spin models possess a classical
counterpart which is unambiguously either integrable or not
\cite{reichl2004, steinigeweg2005}. Thus, in the following we consider a Heisenberg spin chain in an
external magnetic field $B$. Concretely, the Hamiltonian reads
\begin{equation}
\hat{H} = \frac{1}{2} \sum_{\mu=1}^{N} (B + B_\mu) \, \hat{\sigma}_\mu^z
+ \frac{\lambda}{4} \sum_{\mu=1}^{N-1} \hat{\bf \sigma}_\mu \cdot
\hat{\bf \sigma}_{\mu+1} \; ,\; N \; \text{even} , \label{heisenberg}
\end{equation}
where $\hat{\bf \sigma}_\mu$ are the standard Pauli operators and
$B_\mu$ are local variations from $B$. Furthermore, $B_\mu$ are
chosen as Gaussian distributed random numbers with $\langle \,
B_\mu \, \rangle = 0$ and $\langle B_\mu B_\nu \rangle =
\delta_{\mu \nu} \, \epsilon^2$. The parameter $\lambda$ sets the
coupling strength, but the Heisenberg interaction is not
normalized to 1, unlike (\ref{interaction}). As mentioned above we
operationally divide the chain into only two subunits, namely the
first and second half. Note that due to the local fields $B_\mu$
the spectra of those halves may not be identical.
\begin{figure}
  \centering
  \includegraphics[width=14cm]{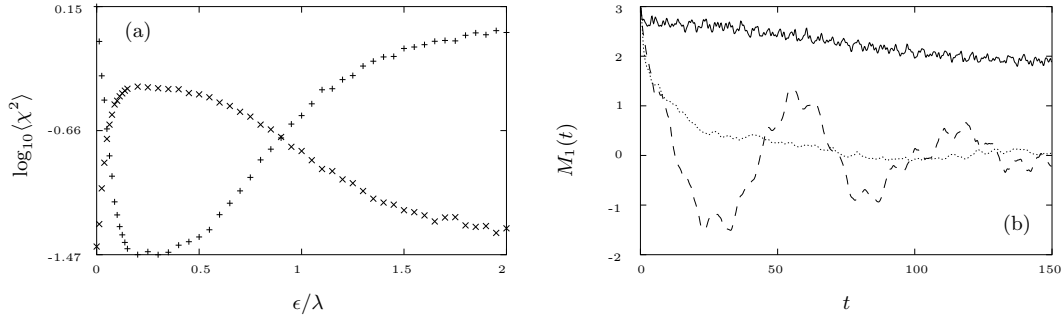}
  \caption{\textbf{(a)} Average deviation of the level statistics from
  the Wigner distribution ($+$) and the Poisson distribution ($\times$)
  for the spin model corresponding to (\ref{heisenberg}). \textbf{(b)}
  Time evolution of the first half's local magnetization for
  $\epsilon / \lambda = 0$ (dashed curve), $0.25$ (dotted curve)
  and $1$ (solid curve). Evidently, exponential decay indicating normal
  transport occurs only for $\epsilon / \lambda = 0.25$, the
  parameter for which the level statistics are most Wigner-like.}
  \label{fig7}
\end{figure}

Obviously, (\ref{heisenberg}) is invariant under rotations around
the $z$-axis. In order to compute the NNLSD we choose the subspace
with $M = 0$, where $M$ is the quantum number with respect to
$\hat{S}^z = \sum_{\mu = 1}^N \hat{\sigma}_\mu ^z/2$, the
generator of these rotations. For a model as given by
(\ref{heisenberg}) with $N = 12$ the dimension of this subspace
is $d = 924$. Figure \ref{fig7}a shows $\chi_W^2$ versus
$\epsilon$, averaged for $100$ sequences $B_\mu$, that is, the
average deviation of the NNLSD from the Wigner distribution, and
$\chi_P^2$, respectively, cf.\ (\ref{chisquare}). Since the system
is integrable for $\epsilon = 0$, a Poisson-like NNLSD is
obtained. When $\epsilon$ increases from zero, the system
undergoes a transition to chaos and consequently the NNLSD becomes
Wigner-like. The minimum of $\chi_W^2$ (the maximum of $\chi_P^2$)
is reached at $\epsilon \approx \lambda / 4$. When $\epsilon$
further increases and becomes larger than $\lambda$, the system
becomes localized and accordingly a Poisson-like NNLSD reappears,
see \cite{avishai2002, kudo2004, santos2004}, too.

Since neither the level statistics nor the dynamics within the
mentioned subspace depend on the constant field $B$, we set $B =
0$. Here we analyze the transport of magnetization which is, just 
like energy, a conserved quantity in this model. According to our
above partition scheme we define two local magnetizations:
\begin{equation}
M_1 := \frac{1}{2} \langle \psi(t) | \sum_{\mu = 1}^6 \hat{\sigma}_\mu ^z | 
\psi(t) \rangle \quad \textnormal{and} \quad M_2 := \frac{1}{2} \langle 
\psi(t) | \sum_{\mu=7}^{12} \hat{\sigma}_\mu ^z | \psi(t) \rangle  \; .
\end{equation}
If the transport behavior of
magnetization was diffusive on this scale, the $M$'s should
exhibit a dynamics as generated by a direct analogue to
(\ref{rates}), i.e., they should simply relax exponentially to
equilibrium. In Fig.\ \ref{fig7}b $M_1$ is displayed for the cases
$\epsilon / \lambda = 0$ (dashed), $0.25$ (dotted) and $1$
(solid). The chosen initial state $\psi(0)$ is the only state with
$M_1 = 3$ and $M_2 = -3$. Obviously, for the case $\epsilon /
\lambda = 0$ the transport is non-normal, the ``bouncing''
behavior of the magnetization could rather be interpreted as a
hint for ballistic transport. This issue has been discussed very
controversially in the literature \cite{narozhny1998, rabson2004, fabricius1998,
zotos2003, benz2006}. So far, we also have no definite conclusion.
For $\epsilon / \lambda = 1$ almost no transport is observable,
the magnetization seems to be stuck, i.e., localized. But at the
minimum of $\chi_W^2$ (the maximum of $\chi_P^2$), $\epsilon /
\lambda = 0.25$, there is a tolerably good agreement with
(\ref{rates}): the first half's local magnetization decays almost
exponentially from the initial value $M_1 = 3$ to the equilibrium
value $M_1 = 0$, as expected for normal transport.

What is to be expected for longer chains? Regular transport
and localization scenarios should be unaffected in the thermodynamic
limit as the investigations of the first system class and theories in 
\cite{michel2005, gemmer2006-1, gemmer2006-2} suggest. The
ballistic-like transport may possibly become normal on a larger scale.

Of course, our Heisenberg model can be mapped on a 1D twelve-site Hubbard
model of interacting spinless fermions with disorder. Such systems
have recently been studdied in \cite{basko2006} with the result that
the 
interaction may lead to finite conductivity (above some critical temperature) for systems that would be
localized (in real space) otherwise. Although our states are far from being
thermal, our results are in accord with those findings: If we remove
the interactions (the $\hat{\sigma}_\mu ^z
\hat{\sigma}_{\mu+1}^z$-term) for the normal transport case ($\epsilon /
\lambda = 0.25$) we find that magnetization essentially stays in the
first half, i.e., localization occurs. Consistently, the NNLSD becomes
Poissonian again.

Let us finally summarize and conclude: We investigated several
finite one-dimensional quantum systems. We introduced definitions
for perfectly diffusive, normal transport behavior and for the
level statistics expected from a perfectly chaotic system (Wigner
statistics). By numerically solving the corresponding
Schr\"odinger equations we got the exact energy spectra as well as
the exact dynamics of our systems. With those data we computed the
deviation of our model's level statistics from a perfectly chaotic
system's level statistics and the deviation of our model's exact
transport dynamics from a perfectly diffusive transport dynamics.
From comparing those deviations we found that models featuring
nearly perfect diffusive transport always feature nearly perfect
chaotic level statistics, while the inverse is not true. This
result is eye-catching and eventually leads to our conclusion that
a Wigner-like NNLSD might be a necessary, but not sufficient
condition for normal transport.

\acknowledgments

We thank H.-J.\ Schmidt   for fruitful discussions
and gratefully acknowledge financial support by the Deutsche
Forschungsgesellschaft.

\end{document}